\documentclass{emulateapj}
\usepackage{amsmath}
\usepackage{graphicx}

\shorttitle{Eccentricity Affects Detection of Transiting Planets}
\shortauthors{Christopher J. Burke}
\slugcomment{Submitted for publication in the Astrophysical Journal}
\received{2007 November 09}
\begin{document}
\title{Impact of Orbital Eccentricity on the Detection of Transiting Extrasolar Planets}
\author{Christopher J. Burke}
\affil{Space Telescope Science Institute, 3700 San Martin Dr., Baltimore, MD, 21218}
\email{cjburke@stsci.edu}

\begin{abstract}
For extrasolar planets with orbital periods, P$>$10 days, radial
velocity surveys find non-circular orbital eccentricities are common,
$\langle e\rangle\sim 0.3$.  Future surveys for extrasolar planets
using the transit technique will also have sensitivity to detect these
longer period planets.  Orbital eccentricity affects the detection of
extrasolar planets using the transit technique in two opposing ways:
an enhancement in the probability for the planet to transit near
pericenter and a reduction in the detectability of the transit due to
a shorter transit duration.  For an eccentricity distribution matching
the currently known extrasolar planets with P$>$10 day, the
probability for the planet to transit is $\sim 1.25$ times higher than
the equivalent circular orbit and the average transit duration is
$\sim 0.88$ times shorter than the equivalent circular orbit.  These
two opposing effects nearly cancel for an idealized field transit
survey with independent photometric measurements that are dominated by
Poisson noise.  The net effect is a modest $\sim 4\%$ increase in the
transiting planet yield compared to assuming all planets have circular
orbits.  When intrinsic variability of the star or correlated
photometric measurements are the dominant source of noise, the transit
detectability is independent of the transit duration.  In this case
the transit yield is $\sim$25\% higher than that predicted under the
assumption of circular orbits.  Since the Kepler search for
Earth-sized planets in the habitable zone of a Solar-type star is
limited by intrinsic variability, the Kepler mission is expected to
have a $\sim$25\% higher planet yield than that predicted for circular
orbits if the Earth-sized planets have an orbital eccentricity
distribution similar to the currently known Jupiter-mass planets.
\end{abstract}


\section{Introduction}

The known extrasolar planets possess a broad distribution of orbital
eccentricity \citep{BUT06} (see Figure~\ref{fig:ecchist}).  The short
period, Hot Jupiter (P$<$10 day) planets predominately have circular
orbits.  However, at longer orbital periods circular orbits become a
minority and the median eccentricity for extrasolar planets
e$\sim$0.3.  At the extreme eccentricity end, there are three planets,
HD 80606b \citep{NAE01}, HD 20782b \citep{JON06}, and HD 4113b
\citep{TAM07}, having e$>$0.9.  HD 80606b comes closer to its stellar
host (a=0.033 AU) than many of the circular orbit Hot Jupiter planets.
\citet{FOR07} recently reviewed the various mechanisms invoked to explain the
distribution of eccentricities for the known extrasolar planets.
Interactions with a stellar companion, planetary companion, passing
star, gaseous disk, planetesimal disk, and stellar jets have all been
proposed to modify the orbital eccentricity of extrasolar planets.

Limited discussions in the literature have been given to the impact
orbital eccentricity has on a transit survey for extrasolar planets.
\citet{TIN05} discuss the impact of eccentricity on their $\eta$
parameter ($\eta$ is the ratio of the observed transit duration to an
estimate of the transit duration).  As expected, they find transits
occur near pericenter (apocenter) are shorter (longer) in duration
than the circular orbit case, and they show that the transit duration
of an eccentric is typically shorter than a circular orbit of the same
period.  However, their discussion was focused on the impact of
orbital eccentricity on their $\eta$ parameter. \citet{MOU06} also
discuss how the transit duration is affected by orbital eccentricity,
but they do not quantify the impact this will have on transit surveys.
Recently, \citet{BAR07} derives the probability for a planet on an
eccentric orbit to transit and conclude that the photometric precision
of current surveys and future surveys, such as Kepler, is insufficient
to determine the orbital eccentricity solely from the light curve.
\citet{BAR07} concludes that without knowledge of the eccentricity
from radial velocity data or independent measurement of the stellar
host radius, the habitability of planets detected with Kepler will
remain unknown.

Neglecting the impact eccentricity has on transit detections is
justified for the current sample of transiting planets given the
predominance of circular orbits for the Hot Jupiters and the strong
bias of transit surveys against finding long period planets on
circular orbits \citep{GAU05}.  The announced transit for the planet
orbiting HD 17156\footnote{First detected by the radial velocity
technique \citep{FIS07}} with a 21 day period and eccentric orbit
\citep{BARB07} is a precursor for the kinds of planets detectable in
transit surveys.  As transit surveys continue, longer period
transiting planets may be discovered.  More importantly, the recently
launched COROT mission will surpass current surveys for sensitivity to
longer period planets \citep[P$\sim$ several months; ][]{BORD03}.
Also, the Kepler mission, scheduled for launch in 2009, has a goal to
find Earth-sized objects at 1 AU from their host star \citep{BORU04}.
The main purpose of this paper is to show that given the distribution
of eccentricities for the currently known extrasolar planets,
eccentricity should not be ignored in assessing the detectability of
transiting giant planets when the transit survey is sensitive to
planets with P$>$ 10 day.

In addition to longer period planets, the COROT and Kepler missions
also will detect transiting planets with small, Earth-sized radii.
The eccentricity distribution for planets less massive than the
currently known Jupiter-mass planets is beginning to be explored.
\citet{RIB07} and \citet{FOR07} provide tentative evidence for the
tendency of lower mass planets to have lower eccentricities in the
current sample of radial velocity planets.  There is theoretical
agreement that a proto-planetary gas disk strongly damps the
eccentricity of non-gap opening embedded low-mass planets
\citep{GOL80, CRE07}.  However, the theoretical models show that once
the gas disk dissipates, dynamical interactions amongst the planets
results in a random-walk diffusion that leads to an increasing
eccentricity that takes on a Rayleigh-distribution similar to what is
observed \citep[dotted line in Figure~\ref{fig:ecchist};
][]{JUR07,ZHO07}.  Opposing the increases in eccentricity from
planet-planet scattering, late stage interactions with planetesimals
can preferentially damp the eccentricities of lower mass planets
enabling theoretical models to achieve the low eccentricities of the
terrestrial planets of the Solar System \citep{RAY06} and possibly
explain the current trend of lower eccentricities for lower mass
planets \citep{FOR07}.  Given the number of potential physical processes that can affect 
orbital eccentricity, it is premature to assume that the typical
Earth-like planet has an eccentricity near zero like the Solar System.

In this study, \S~\ref{sec:eccdist} reviews the observed eccentricity
distribution of the known radial velocity extrasolar planets.  The
broad distribution of orbital eccentricity has two main effects on the
sensitivity of a transit survey.  First, the planet-host separation
varies along an eccentric orbit, enhancing the probability to transit
when the planet is relatively closer to the stellar host.
\S~\ref{sec:tranprob} quantifies the net affect on the transit
probability for a population of planets with non-circular orbits.
Second, the planet velocity varies along and eccentric orbit,
resulting in a reduction or lengthening of the transit duration.
\S~\ref{sec:trandur} quantifies the distribution of transit durations
resulting from a population of planets on eccentric orbits.  
\S~\ref{sec:disc} describes how the transit duration affects transit
detection for transit surveys in the limit of various noise sources.
\S~\ref{sec:conc} concludes by quantifying the net result of the two
aforementioned effects, the enhanced probability to transit and the
reduced detectability, on the yield from transit surveys.

\section{Eccentricity Distribution}\label{sec:eccdist}

\begin{figure}
\includegraphics[scale=0.5]{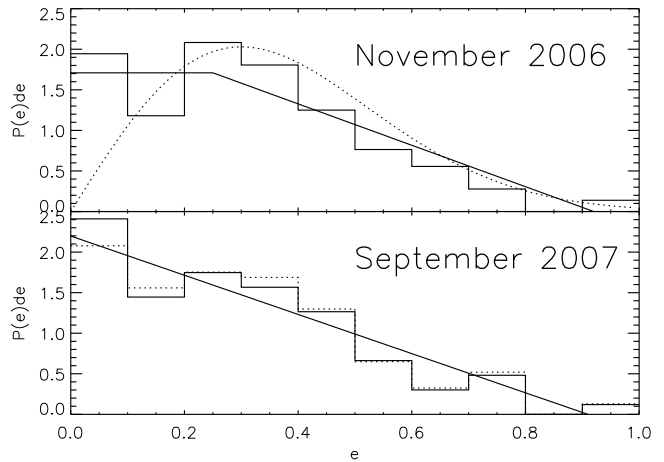}
\caption{Normalized histogram of the eccentricity distribution for known extrasolar planets with P$>$10 day.  The eccentricity distribution is given for the known extrasolar planets at two epochs: November 2006 ({\it Top Panel}) and September 2007 ({\it Bottom Panel}).  For this study a two-piece model parameterizes the eccentricity distribution ({\it solid line}).  The model has 2 parameters: $e_{\rm crit}$ and $e_{\rm max}$.  The model distribution is uniform at low eccentricity up to $e_{\rm crit}$; and toward higher eccentricity, the model linearly decreases to zero-value at $e_{\rm max}$.  Also shown in the top panel is the Rayleigh distribution that describes the eccentricity distribution from the planet-planet scattering calculations of \citep{JUR07} ({\it dotted line}).  The dotted histogram in the bottom panel shows the eccentricity distribution from the latest epoch (September 2007) that excludes planets with P$<$ 20 day.\label{fig:ecchist}}
\end{figure}

The solid histograms in Figure~\ref{fig:ecchist} show the normalized
distribution of eccentricities for the known extrasolar planets with
$P>10$ day \citep{BUT06}.  The top panel shows the eccentricity
distribution from planets reported before November 2006 and the bottom
panel shows the eccentricity distribution including more recent
discoveries (September 2007).  The more recent radial velocity
discoveries over the last year have increased the sample of low
eccentricity systems.  Results from this study are given for both
epochs of the observed eccentricity distribution in order to
characterize how the uncertainty in the underlying orbital
eccentricity distribution affects the results.  No attempt was
made to select planets with well determined eccentricities, avoidance
of multiple planet or multiple star systems, correction for observational
biases against high eccentricity planets
\citep{CUM04}, or evolution of orbital elements due to tidal
circularization.  Planets with $P<10$ day are not included in order to
examine the potential issues that future transit surveys sensitive to
longer period planets will encounter.  In this period regime (P$>$10
day) high eccentricity planets are quite common ($\langle
e\rangle=0.3$ and $\sigma_{e}=0.2$).

A two-piece model parameterizes the eccentricity distribution from the
earlier epoch (top panel of Figure~\ref{fig:ecchist}).  The first
piece at low eccentricity is flat up to $e_{\rm crit}$.  The second piece matches the first piece at $e_{\rm crit}$, linearly decreases toward
higher eccentricities, and is zero at $e_{\rm max}$.  The normalized two-piece model is given by the equation,

\begin{equation}
P(e)de=\left\{ \begin{array}{ll}
 \frac{2}{e_{\rm crit}+e_{\rm max}}               & 0 \le e \le e_{\rm crit} \\
 \frac{2}{e_{\rm crit}+e_{\rm max}}\frac{(e-e_{\rm max})}{(e_{\rm crit}-e_{\rm max})} & e_{\rm crit} < e \le e_{\rm max} \\
    0                                          & e_{\rm max} < e < 1  \end{array}
       \right. .\label{eq:eccdist}
\end{equation}

A $\chi^{2}$ minimization yields the best model parameters $e_{crit}=0.25$ and
$e_{max}=0.92$.  The $\chi^{2}$ minimization applied to the more recent
epoch eccentricity distribution (bottom panel of Figure~\ref{fig:ecchist}),
yields $e_{crit}=0.0$ and $e_{max}=0.91$.  The relative
increase in the number of low eccentricity planets discovered in the
last year results in best parameters that are effectively a single-piece model.

\citet{JUR07} and \citet{ZHO07} predict that the eccentricity distribution of dynamically active planetary
systems approaches a Rayleigh-distribution due to planet-planet
scattering, and they show the Rayleigh-distribution is similar to the
observed planet eccentricity distribution.  The dotted curve in
Figure~\ref{fig:ecchist} shows the Rayleigh-distribution with
$\sigma_{e}=0.3$ that best describes the outcome of the planet-planet
scattering calculations from
\citet{JUR07}.  The Rayleigh-distribution under represents the
low eccentricity systems in the most recent eccentricity data.
\citet{JUR07} based their study on the sample of planets known at the time (April 2006).  Subsequently, relatively more planets
with low eccentricities have been announced.  In addition,
\citet{JUR07} included systems with P$>$20 day
whereas this sample has P$>$10 day.  The dotted histogram in
the lower panel of Figure~\ref{fig:ecchist} shows the eccentricity
distribution at the most recent epoch (September 2007), but with
systems with P$<$20~day removed; apparently the difference is slight
and will hereafter be ignored.  Although this study concentrates on
the two-piece model to describe the eccentricity
distribution, results are also given for the
Rayleigh-distribution with $\sigma_{e}=0.3$.  To provide results with
the Rayleigh-distribution, the distribution is set to zero for $e>0.95$.

\section{Transit Probability}\label{sec:tranprob}

\citet{BAR07} derives the impact of orbital eccentricity on the
probability for a planet to transit its stellar host \citep[see
also][]{SEA03}.  The probability to transit depends on the planet-star
separation during transit.  For an eccentric orbit, a transit occurring
during pericenter enhances the transit probability and a transit
occurring during apocenter decreases the transit probability.  When
averaged over observing angles, the net result is an enhancement of
the probability to transit over the circular orbit case,
\begin{equation}
{\rm Prob}_{{\rm T}e}=\frac{{\rm Prob}_{{\rm T}o}}{(1-e^{2})},
\end{equation}
where ${\rm Prob}_{{\rm T}o}$ is the transit probability of the
circular orbit case \citep{BAR07}.  A planet with $e$=0.6 is $\sim$1.5
times more likely to transit than a planet on a circular orbit.

\begin{figure}
\includegraphics[scale=0.9,viewport=0 100 350 350]{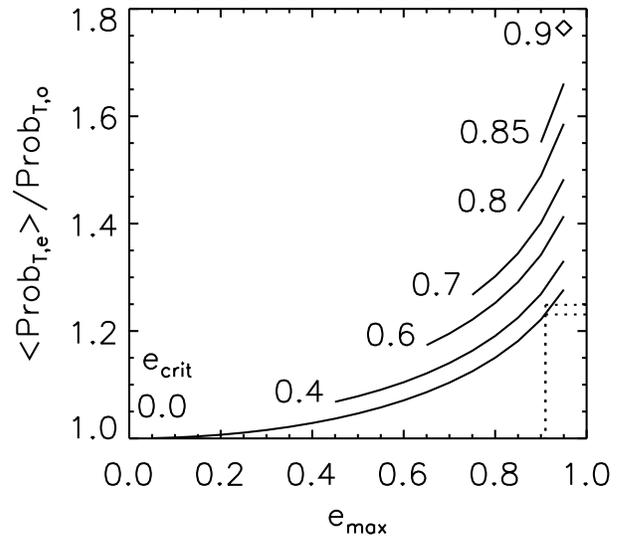}
\caption{Average transit probability as a function of the observed eccentricity distribution model parameters.  The transit probability is scaled to the equivalent transit probability for a circular orbit.  The abscissa indicates $e_{\rm max}$, and the curves are for selected values of $e_{\rm crit}$ as
labeled.  For the model parameters shown in Figure~\ref{fig:ecchist}, $\langle{\rm Prob}_{{\rm T}e}\rangle\sim 25\%$ higher than assuming all planets have circular orbits ({\it dotted lines}).\label{fig:probtran}}
\end{figure}

With
the observed eccentricity distribution from \S~\ref{sec:eccdist}, it
is possible to calculate the average enhancement in the transit
probability over the circular orbit case,
\begin{equation}
\langle {\rm Prob}_{{\rm T}e}\rangle=\int^{1}_{0} {\rm Prob}_{{\rm T}e}P(e)de.
\end{equation}
The resulting integral using Equation~\ref{eq:eccdist} for the eccentricity distribution is,
{\scriptsize
\begin{equation}
\begin{split}
 & \langle {\rm Prob}_{{\rm T}e}\rangle=\frac{{\rm Prob}_{{\rm T}o}}{(e_{\rm max}^2-e_{\rm crit}^2)}\left[2(e_{\rm max}-e_{\rm crit}){\rm arctanh}(e_{\rm crit})\right. \\
 & \left. +\ln\left( \frac{(1-e_{\rm max}^2)}{(1-e_{\rm crit}^2)}\right) +e_{\rm max}\ln\left(\frac{(1-e_{\rm crit})(1+e_{\rm max})}{(1+e_{\rm crit})(1-e_{\rm max})}\right) \right].
\end{split}
\label{eq:probtrane}
\end{equation}}
Using the numerical values for $e_{\rm crit}$ and $e_{\rm max}$ from
\S~\ref{sec:eccdist}, $\langle {\rm Prob}_{{\rm T}e}\rangle$=1.26 and
1.23 for the earlier and current epochs of observed eccentricities,
respectively.  The Rayleigh distribution results in $\langle {\rm
Prob}_{{\rm T}e}\rangle$=1.31.  Figure~\ref{fig:probtran} illustrates
$\langle {\rm Prob}_{{\rm T}e}\rangle$ for other choices of the
eccentricity distribution model parameters.  Figure~\ref{fig:probtran}
is a function of $e_{\rm max}$ and the curves are for selected values
of $e_{\rm crit}$ as labeled.  Thus, it is expected that the yield
from a transit survey that is sensitive to P$>$10 day planets will be
$\sim$25\% larger than expectations that assume circular orbits for
all planets.  This section discusses only the probability to transit,
and it does not address whether the planet has a transit signal that
is detectable.  Thus, the results from this section assumes the
enhanced probability for the planet to transit does not affect its
detectability, which is the topic of the next section.

\section{Transit Duration}\label{sec:trandur}

\subsection{Edge-on Transit}\label{sec:i90}
In addition to enhancing the probability for a transit to occur,
orbital eccentricity results in the transit duration varying according
to the planet's longitude of pericenter during transit.  The
transit duration is shortest (longest) if the transit occurs when the
planet is at pericenter (apocenter) when transiting its stellar host.
A short transit duration may reduce the detectability of the transit
event.  Assuming constant velocity during transit, the transit
duration scaled to the edge-on ($i=90\degr$) circular orbit case has
extrema of
\begin{equation}
\tau_{\frac{p}{a}}=\frac{t_{\frac{p}{a}}}{t_{o}}=\sqrt{\frac{(1\mp e)}{(1\pm e)}}\sim  1\mp e+e^2/2,\label{eq:tranextrema}
\end{equation}
where the sign on top and bottom corresponds to the pericenter and
apocenter transit durations, respectively \citep{BAR07}.  When
$e$=0.6, the transit duration at apocenter is twice the circular orbit
case.  Given the range of eccentricities observed for extrasolar
planets with P$>$10 day, we expect significant variations in the
transit duration compared to that of a circular orbit.

\citet{TIN05} (their Equation 7) provides a simplified form of the
transit duration, including longitude of pericenter, $\varpi$, and
orbital inclination, ${\it i}$.  In deriving the transit duration,
\citep{TIN05} assume constant orbital velocity and planet-star
separation during transit and the planet crosses the stellar disk
along a straight path ($a\gg R_{\star}$).  An exact calculation (of a
numerical nature given the need to solve Kepler's Equation) of the
transit duration finds Equation 7 of
\citet{TIN05} is accurate to better than 5\% for
$e$=0.9 and orbital separation $a\geq 0.1$ AU (P$\sim$10 day).  The
equation is accurate to better than 1\% for $e$=0.9 and $a\geq 1.0$
AU.

After scaling to the transit duration of the edge-on ($i=90\degr$), circular orbit with the same period and separation, Equation 7 of \citet{TIN05} becomes,
\begin{equation}
\tau=\frac{\sqrt{(1-e^2)}}{(1+e\cos(\varpi))}\sqrt{1-\rho^{2}\left[ \frac{(1-e^{2})}{(1+e\cos(\varpi))}\right] ^{2}\cos(i)^{2}},
\label{eq:durateq}
\end{equation}
where $\rho=a/(R_{\star}+R_{p})$, $R_{\star}$ is the radius of the star and $R_{p}$ is the radius of the planet.  The definition for $\varpi$ is with respect to the line of sight.  Thus, $\varpi=0$ means the pericenter is aligned with the observers line of sight, and $\varpi=180\degr$ means the apocenter is aligned with the observers line of sight.  This varies from the definition of the argument of pericenter, $\omega$, which is defined with respect to the line of nodes on the plane of the sky ($\varpi=\omega-90\degr$).

\begin{figure}
\includegraphics[scale=0.5,viewport=0 100 350 350]{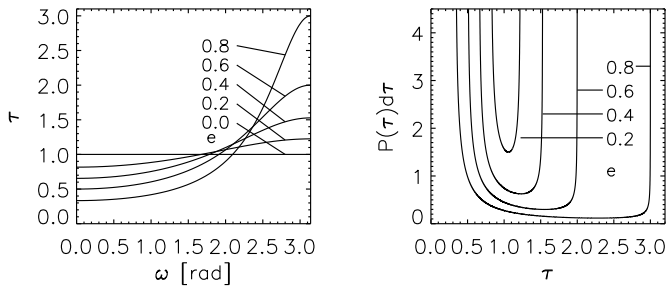}
\caption{{\it Left}: Edge-on transit duration, $\tau$, scaled to the edge-on circular orbit case as a function of the longitude of pericenter, $\varpi$, for several eccentricities as labeled.  {\it Right}: For a uniform distribution of $\varpi$, solid curves show the distribution for transit duration scaled to the edge-on circular orbit case assuming all orbits are edge-on (i.e. $i=90\degr$).\label{fig:duratboth}}
\end{figure}

The $i=90\degr$ case illustrates the first order impact of orbital eccentricity on transit duration.  In this case, Equation~\ref{eq:durateq} simplifies to
\begin{equation}
\tau_{90\degr}=\frac{\sqrt{(1-e^2)}}{(1+e\cos(\varpi))}.
\label{eq:trandurati90}
\end{equation}
The left panel of Figure~\ref{fig:duratboth} shows the transit
duration with respect to the edge-on circular orbit case for a variety of
orbital eccentricities as a function of $\varpi$.  For a uniform
distribution in $\varpi$, the shallow slope in transit
duration at pericenter and apocenter results in a distribution of
transit durations peaked at these extrema (given by Equation~\ref{eq:tranextrema}) with low probability for a
transit duration in between these extrema.

Assuming a uniform distribution of $\varpi$, $p(\varpi)d\varpi=d\varpi/\pi$, the transformation law of probabilities \citep[Equation 7.2.4 in][]{PRE92} yields the distribution of transit durations,
\begin{equation}
p(\tau)d\tau=\left| \frac{\partial \varpi}{\partial \tau} \right| p(\varpi)d\tau, 
\end{equation}
resulting in  
\begin{equation}
p(\tau)d\tau = \frac{\sqrt{1-e^{2}}d\tau }{\pi \tau\sqrt{(e\tau)^{2}-(\sqrt{1-e^{2}}-\tau )^{2}}}.
\end{equation}
The right panel of Figure~\ref{fig:duratboth} shows the probability
density for transit duration with $i=90\degr$ for various orbital
eccentricities.  As expected the probability density is heavily
weighted toward the extrema.  The singularities in the probability
density at the extrema are integrable, making the probability density
normalizable.

\subsection{Affect of Varying Inclination Angle}\label{incdist}

The previous section shows the expected transit duration for fixed
orbital eccentricity when the inclination angle, $i=90\degr$.  Varying
the inclination impacts the transit duration in two important ways.
First, as the inclination decreases, the path of the planet across the
stellar disc shortens, resulting in a reduction of the transit
duration.  Second, for an eccentric orbit, the planet is closer to the
star at pericenter and farther away at apocenter.  The smaller
separation at pericenter increases the range of inclination angles
that result in a transit, and conversely, a planet at apocenter will
have a reduction in the range of inclination angles for a transit to
occur.  Thus, long transit duration events when the planet is at
apocenter become increasingly rare as the eccentricity increases.
This section quantifies these effects on the transit duration
distribution.

Beginning with a simple joint distribution that is uniform in $\cos(i)$
and uniform in $\varpi$ enables the more complicated
distribution for transit duration to be derived.  For this section, $e$ remains
fixed.  The beginning joint distribution is,
\begin{equation}
P(\varpi,0 \le \cos(i) \le \cos(i_{min})| e)d(\cos(i))d\varpi=Ad(\cos(i))d\varpi,
\label{eq:cosweven}
\end{equation}
where $\cos(i_{min})$ is cosine of the minimum inclination angle
necessary for a transit to occur, and $A$ is the normalization
constant.  Setting $\tau=0$ in Equation~\ref{eq:durateq} and solving
for $\cos(i)$ yields $\cos(i_{min})=(1+e\cos(\varpi))/\rho (1-e^{2})$.
Integrating Equation~\ref{eq:cosweven} over the range of variables
yields the normalization constant, $A=\rho(1-e^{2})/\pi$.

The transformation law of probabilities for multiple dimensions
\citep[Equation 7.2.4 in][]{PRE92} provides the joint probability
density in terms of new variables $\tau$ and $\varpi$ when
starting with the probability density for $\cos(i)$ and
$\varpi\prime$.  
\begin{equation}
P(\tau,\varpi | e)d\tau d\varpi=\left\| \begin{array}{cc}
                         \frac{\partial \cos(i)}{\partial \tau} & \frac{\partial \cos(i)}{\partial \varpi} \\
                         \frac{\partial \varpi\prime}{\partial \tau} & \frac{\partial \varpi\prime}{\partial \varpi} \end{array}\right\|
P(\cos(i),\varpi\prime)d\tau d\varpi,
\label{eq:ptauomega}
\end{equation}
In Equation~\ref{eq:ptauomega}, $\partial \varpi / \partial \tau=0$ and $\partial \varpi\prime / \partial \varpi =1$.  Thus, only $\partial \cos(i) / \partial \tau$ remains, and Equation~\ref{eq:ptauomega} simplifies to
\begin{equation}
P(\tau ,\varpi |e)d\tau d\varpi=\left| \frac{\partial \cos(i)}{\partial \tau} \right| \rho (1-e^{2})/\pi d\tau d\varpi.
\label{ptauomega2}
\end{equation}

The requisite derivative needed in Equation~\ref{ptauomega2} is
obtained by solving Equation~\ref{eq:durateq} for $\cos(i)$ yielding
\begin{equation}
\cos(i)=\frac{(1+e\cos(\varpi))}{\rho(1-e^2)^{3/2}}\sqrt{(1-e^2)-\tau^{2}(1+e\cos(\varpi))^{2}},
\label{cosi}
\end{equation}
and taking
the partial derivative with respect to $\tau$.  Performing this operation yields
\begin{equation}
\begin{split}
 & P(0\leq \tau \leq \tau_{a}, \varpi_{\rm min}\leq \varpi \leq \pi|e)d\tau d\varpi \\
 & =\frac{\tau (1+e\cos(\varpi))^{3}}{\pi \sqrt{1-e^{2}} \sqrt{(1-e^{2})-\tau^{2}(1+e\cos(\varpi))^{2}}} d\tau d\varpi , 
\end{split}
\label{eq:ptauomega3}
\end{equation}
where the lower limit to $\varpi$ is necessary to avoid an imaginary result and is given by 
\begin{equation}
\varpi_{\rm min}=\cos^{-1}\left[ {\rm MIN}\left( 1.0,\frac{\sqrt{1-e^{2}}-\tau}{\tau e} \right) \right].
\end{equation}
For $\tau \leq \tau_{peri}$, $\varpi_{\rm min}=0$, but as $\tau>\tau_{peri}$,
$\varpi_{\rm min}> 0$ since only transits occurring closer to apocenter
result in a transit duration long enough.

Finally, the conditional
probability density for transit duration alone is obtained by
integrating over $\varpi$,
\begin{equation}
P(\tau | e) d\tau = \int^{\pi}_{\varpi_{\rm min}} P(\tau ,\varpi |e)d\tau d\varpi.
\label{eq:ptauconde}
\end{equation}
A solution to the above integral is possible in terms of a summation
of the incomplete elliptic integrals of the first, second, and third
kind \citep{BYR54}.  In practice, I choose to solve the integral
numerically, which is readily solved using the
Romberg open ended algorithm which takes into account the singularity
at the lower limit of integration at $\varpi=\varpi_{\rm min}$ \citep{PRE92}.
Tests of convergence show Equation~\ref{eq:ptauconde} has a singularity at
$\tau=\tau_{peri}$, but it is integrable elsewhere over the range $0\leq \tau \leq
\tau_{a}$.

\begin{figure}
\includegraphics[scale=0.5,viewport=0 100 350 350]{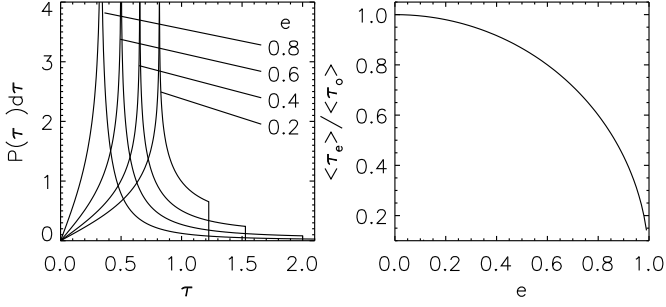}
\caption{{\it Left:} Transit duration distribution scaled to the edge-on circular orbit case for a uniform distribution of $\cos(i)$ and fixed eccentricity.  The cases $e$=0.2, 0.4, 0.6, and 0.8 are shown.  {\it Right:} Average transit duration scaled to the average transit duration of a circular orbit as a function of orbital eccentricity ({\it solid line}).\label{fig:duratinc}}
\end{figure}

The left panel in Figure~\ref{fig:duratinc} shows the probability
density for transit duration scaled to the edge-on circular orbit case
at several values of eccentricity.  By comparison to
Figure~\ref{fig:duratboth}, the main impact of orbital inclination is to
strongly enhance the probability of observing a short duration event
at pericenter relative to a long duration event at apocenter.  The
probability density also allows arbitrarily short events due to the
potential for grazing events.

The probability density for transit duration can be summarized by finding the average transit duration at fixed eccentricity,
\begin{equation}
\langle \tau_{e}\rangle =\int^{\tau_{a}}_{0}\tau P(\tau | e) d\tau.\label{eq:avgtau}
\end{equation}
The right panel in Figure~\ref{fig:duratinc} shows $\langle \tau_{e}\rangle $ scaled
to the average transit duration of the circular orbit case,
$\langle \tau_{0}\rangle =\pi /4$, as the solid line.  The function,
\begin{equation}
\langle\tau_{e}\rangle=\frac{\pi}{4}\sqrt{1-e^2},
\end{equation}
fits the relation to better than $10^{-6}$ (similar to the numerical
integration precision), which very strongly suggests this is the
analytical solution to the integral in Equation~\ref{eq:avgtau}.
\citet{TIN05} find the same result in terms of the average value for
their $\eta$ parameter at fixed eccentricity (see their Equation 18).

\subsection{Transit Duration Distribution For Observed Eccentricity Distribution}\label{sec:tranprobecc}

The previous section derives the transit duration
distribution at fixed eccentricity.  This section derives the
transit duration distribution assuming planets follow the
observed eccentricity distribution.  As in the previous section, 
deriving the transit duration distribution begins with the simple
distribution that is uniform in $\cos(i)$, uniform in $\varpi$, and
the distribution of $e$ follows the observed distribution as given in
\S\ref{sec:eccdist}.  The transformation law of probabilities enables
transforming the simple distribution in $\cos(i)$, $\varpi\prime$, and
$e\prime$ into the distribution expressed in $\tau$, $\varpi$, and
$e$.
 
The initial distribution is given by,
\begin{equation}
\begin{split}
 & P(0\le \cos(i) \le \cos(i_{min}), 0 \le \varpi\prime \le \pi , 0 \le e\prime \le e_{max}) \\
 & d\cos(i) d\varpi\prime de\prime =AP(e\prime)d(\cos(i))d\varpi\prime de\prime ,
\end{split}
\end{equation}
where the normalization constant $A=\rho\gamma/\pi$, where 
\begin{equation}
\begin{split}
 & \gamma=(e_{\rm max}^2-e_{\rm crit}^2)\left[2(e_{\rm max}-e_{\rm crit}){\rm arctanh}(e_{\rm crit})\right. \\
 & \left. +\ln\left( \frac{(1-e_{\rm max}^2)}{(1-e_{\rm crit}^2)}\right) +e_{\rm max}\ln\left(\frac{(1-e_{\rm crit})(1+e_{\rm max})}{(1+e_{\rm crit})(1-e_{\rm max})}\right) \right]^{-1}.
\end{split}
\end{equation}
The Jacobian transformation matrix simplifies as before, such that the transit duration distribution scaled to the edge-on, circular orbit transit duration is given by,
\begin{equation}
\begin{split}
 & P(0 \le \tau \le \tau_{a}, \varpi_{\rm min} \le \varpi \le \pi , e_{\rm min} \le e \le e_{\rm max})d\tau d\varpi de \\
 & =\left| \frac{\partial \cos(i)}{\partial \tau}\right| P(\cos(i),\varpi\prime ,e\prime)d\tau d\varpi de,
\end{split}
\end{equation}
where the lower limit to the eccentricity, $e_{min}$, becomes necessary for $\tau > 1$, when too small of an eccentricity cannot produce a transit duration as long as $\tau$.  Solving $\tau_{a}$ for $e$ yields $e_{\rm min}={\rm MAX}[0.0,(\tau^2-1)/(\tau^2+1)]$.  Overall, the joint distribution is given by
\begin{equation}
\begin{split}
 & P(\tau ,\varpi ,e)d\tau d\varpi de= \\
 & \frac{\gamma P(e)}{\pi}\frac{\tau (1+e\cos(\varpi))^{3}}{(1-e^{2})^{3/2}\sqrt{1-e^{2}-\tau^{2}(1+e\cos(\varpi))^{2}}}d\tau d\varpi de.
\end{split}
\end{equation}

Integrating over $\varpi$ and $e$ provides the final probability
density for $\tau$ for the assumed distribution of orbital
eccentricities.  Given the additional complication of integration over
two variables, an analytical solution was not forthcoming.  As in
\S~\ref{incdist}, the singularities in the integrand are integrable,
and the Romberg open ended algorithm which takes into
account the singularity at the lower limits of
integration \citep{PRE92} provides the solution.

\begin{figure}
\includegraphics[scale=0.5,viewport=0 100 350 350]{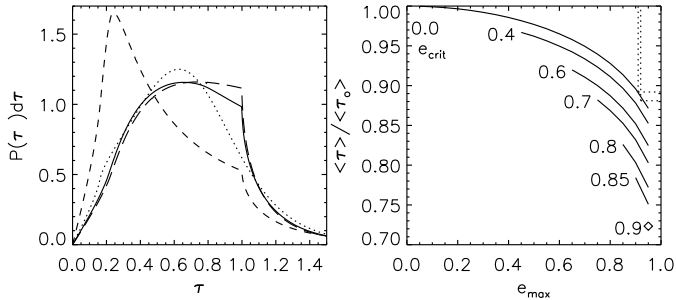}
\caption{{\it Left}: Transit duration distribution scaled to the edge-on circular orbit case for a uniform distribution of $\cos(i)$ and the observed distribution of $e$ as shown in the top panel ({\it solid line}) and bottom panel ({\it long dashed line}) of Figure~\ref{fig:ecchist}.  Given the bias against detecting high eccentricity planets in radial velocity surveys, a uniform distribution of orbital eccentricities up to high eccentricity ($e_{\rm crit}=0.9$ and $e_{\rm max}=0.95$) will be highly skewed towards $\tau\sim 0.3$ ({\it short dashed line}).  A Rayleigh distribution of orbital eccentricities results in relatively fewer transit durations $\tau\sim 1$ due to the relatively fewer objects on circular orbits ({\rm dotted line}).  {\it Right:} Average transit duration scaled to the average transit duration of a circular orbit as a function of the observed eccentricity distribution model parameters.    The abscissa indicates $e_{\rm max}$, and the curves are for selected values of $e_{\rm crit}$ as labeled.  For the model parameters shown in Figure~\ref{fig:ecchist}, $\langle\tau\rangle\sim 0.88$ times shorter than assuming all planets have circular orbits ({\it dotted line}).\label{fig:duratecc}}
\end{figure}

The solid and long-dashed lines in Figure~\ref{fig:duratecc} shows the
distribution of transit duration scaled to the edge-on circular orbit
case for a population of extrasolar planets that follows the observed
orbital eccentricity distributions shown in top and bottom panels of
Figure~\ref{fig:ecchist}, respectively.  The probability density has a
broad flat top with $\tau=0.4$ just as likely to occur as $\tau=1.0$.
For comparison, the transit duration distribution that would result
from a nearly uniform distribution of orbital eccentricity up to high
eccentricities ($e_{\rm crit}=0.9$ and $e_{\rm max}=0.95$) is shown as
the short-dashed line in Figure~\ref{fig:duratecc}.    Given the bias against
detecting $e>0.6$ extrasolar planets in radial velocity studies
\citep{CUM04}, a large population of $e\sim 0.9$ planets cannot be
ruled out.  The transit duration distribution resulting from the Rayleigh
distribution of orbital eccentricity (dotted line in
Figure~\ref{fig:duratecc}) has relatively fewer planets with $\tau\sim
1$ due to its relatively fewer circular orbits.

In the future, if a statistically large sample of transiting planets
with orbital $P>10$ days is available with accurate stellar
parameters, histograms of observed transit duration scaled to the
edge-on circular orbit case may help characterize the underlying
eccentricity distribution.  After accounting for the selection
effects, a large number of $\tau\sim 0.3$ detections relative to
$\tau\sim 0.8$, as illustrated in the right panel of
Figure~\ref{fig:duratecc}, would indicate $e=0.9$ planets are as
common as circular orbits.  Work toward understanding the sensitivity
of Kepler for constraining the underlying eccentricity distribution is
underway (E. Ford, private communication).

The right panel of Figure~\ref{fig:duratecc} summarizes the transit
duration distribution by showing $\langle \tau_{e}\rangle $ in terms of the
parameters characterizing the eccentricity distribution, $e_{\rm
crit}$ and $e_{\rm max}$.  Each line corresponds to a fixed value of
$e_{\rm crit}$ as labeled, and the abscissa indicates $e_{\rm
max}$ of the eccentricity distribution.  For the eccentricity
distribution in the top panel of Figure~\ref{fig:ecchist}
$\langle \tau_{e}\rangle /\langle \tau_{o}\rangle =0.88$ and the bottom panel
$\langle \tau_{e}\rangle /\langle \tau_{o}\rangle =0.89$.  The Rayleigh distribution results in $\langle \tau_{e}\rangle /\langle \tau_{o}\rangle =0.86$

\section{Discussion: Application to Transit Surveys}\label{sec:disc}

The results from \S~\ref{sec:trandur} quantify the impact orbital
eccentricity has on the transit duration.  Overall, orbital
eccentricity results in shorter transit durations than the circular
orbit case, and the short transit duration reduces the
transit detectability.  This section quantifies the reduction in
transit detectability for various noise models of transit
surveys.

In a transit survey with independent photometric measurements, the transit signal to noise ratio is,
\begin{equation}
{\rm SNR}=\frac{\Delta F}{\sigma}\sqrt{N_{\rm obs}},
\end{equation}
where $\Delta F$ is the transit depth (the transit is modeled as a
box-car shape), $\sigma$ is the error on a photometric measurement,
and $N_{\rm obs}$ is the number of measurements during one or more
transit(s).  A shorter transit duration reduces $N_{\rm
obs}\propto\tau$, thus, ${\rm SNR}\propto\sqrt{\tau}$.  The observed
eccentricity distribution (Figure~\ref{fig:ecchist}) results in
$\tilde{\tau}=\langle
\tau_{e}\rangle /\langle \tau_{o}\rangle =0.88$, and on average the
reduced transit duration results in an effectively smaller ${\rm
SNR_{eff}}=\sqrt{\tilde{\tau}}=0.94$ per transit than assuming all
planets are on circular orbits.

The above impact on the transit signal ${\rm SNR}$ due to a shorter
transit duration is for an individual star in a survey.  However, the
reduced ${\rm SNR}$ will have a larger impact on the overall transit
detectability in an ideal field transit survey.  As described in
\citet{GAU05} and \citet{GAU07}, a specified ${\rm SNR}$ criteria for
transit detection, ${\rm SNR_{\rm min}}$, in a field transit survey
corresponds to a maximum distance, $\ell_{\rm max}\propto {\rm
SNR}_{\rm min}^{-1}$, out to which a planet is detectable.  This
proportionality assumes white noise and the dominant source of
photometric error is Poisson noise.  In the studies of \citet{GAU05}
and \citet{GAU07}, $\ell_{\rm max}$ is a function of the planet radius
and stellar host spectral type (i.e. $\ell_{\rm max}$ is a smaller
distance for a smaller radius planet or larger radius star).  For this study only the dependence of $\ell_{\rm max}$ on
transit duration is of interest.

Overall, the yield from a transit survey is
proportional to the number of objects in the survey that meets ${\rm
SNR_{\rm min}}$, which is $N_{\rm obj}\propto
\ell_{\rm max}^{3}$ for stars distributed uniformly in the survey
volume as appropriate for nearby stars.  The effective ${\rm
SNR}_{\rm min,eff}={\rm SNR}_{\rm min}\sqrt{\tilde{\tau}}$ larger than
assuming all planets are on circular orbits.  Thus, the number of
objects in an idealized transit survey where a transit is detectable
is $N_{\rm obj,e}=\tilde{\tau}^{3/2}N_{\rm obj,o}=0.82N_{\rm obj,o}$
times smaller than the case where the detectability of a transit is
based on assuming all planets have circular orbits, $N_{\rm obj,o}$.
Despite the reduced detectability of transits, this is offset by the
higher probability for the planet to transit in the case of
significant eccentricity
\S~\ref{sec:tranprob}.  The overall yield of the idealized transit
survey is discussed in \S~\ref{sec:conc} taking both the reduced
detectability and enhanced probability to transit into account.

In practice, transit surveys typically are affected by correlated
measurements \citep{PON06}.  In this regime, the correlation time
scale is similar to the transit duration and repeated measurements do
not add independent information.  When correlated measurements
dominate the photometric error, the ${\rm SNR}=(\Delta F/\sigma_{\rm
cor})\sqrt{N_{\rm tr}}$, where $N_{\rm tr}$ is the number of transits
detected, and the correlated measurement error, $\sigma_{\rm cor}$, no
longer depends on the stellar luminosity, but is constant
\citep{GAU07}.  When dominated by correlated measurements, the ${\rm
SNR}$ is independent of $\tau$ and the only requirement is to have a
short enough sampling cadence to detect the shortest transit duration
expected.  Thus, a shorter $\tau$ due to orbital eccentricity has no
impact on the transit detectability in a survey which is dominated by
correlated measurements (i.e. $N_{\rm obj,e}=N_{\rm obj,o}$).
However, the Poisson limited, white noise transit survey will have a
higher planet yield than a survey dominated by correlated measurements
\citep{PON06}, but in the latter case, orbital eccentricity does not
impose any additional reduction in the transit detectability.

The Kepler space-based transit survey whose goal is to detect several
Earth-sized planets \citep{BORU04} contends with stellar intrinsic
variability as the dominant source of noise \citep{JEN02}.  Using
integrated flux measurements of the Sun to model intrinsic
variability of stars, \citet{JEN02} characterizes the detectability of
Earth-sized transiting planets with Kepler.  The Sun has low noise on
7 hr time scales typical of transiting Earth-sized planets in 1 yr
orbital periods.  However, the solar noise increases by $\sim$ 4
orders of magnitude by 10 day time scales \citep[see Figure 2 of
][]{JEN02}.  The rapidly increasing intrinsic noise toward longer time
scales cancels any benefit of increased transit signal ${\rm SNR}$ for
transit durations $>$4 hr for the Kepler mission \citep[see Figure 9
of ][]{JEN02}.  Thus, as long as the transit duration remains above 4
hr, orbital eccentricity will not reduce the detectability of
Earth-sized planets with Kepler.

\section{Conclusion}\label{sec:conc}

\begin{figure}
\includegraphics[scale=0.9,viewport=0 100 350 350]{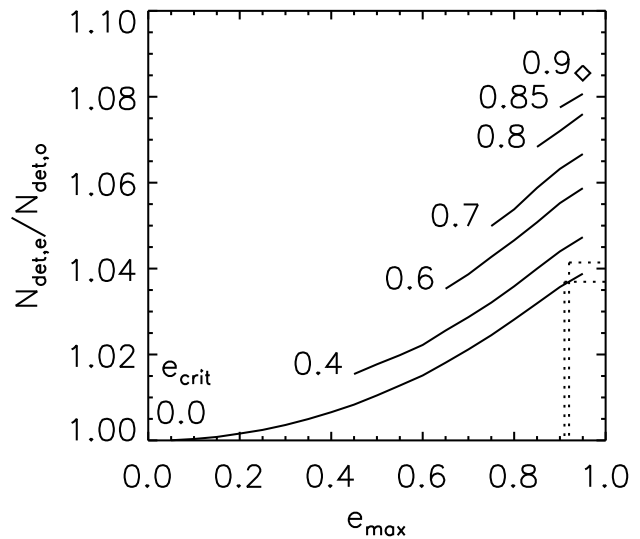}
\caption{The overall yield from an idealized transit survey as function of the observe eccentricity distribution model parameters scaled to the assumption that all planets have circular orbits.  The transit survey is assumed to have independent photometric measurements that are dominated by Poisson
noise.  The abscissa indicates $e_{\rm max}$, and the curves are for selected values of $e_{\rm crit}$ as labeled.  For the model parameters shown in Figure~\ref{fig:ecchist}, $N_{\rm det}\sim 4\%$ higher than assuming all planets have circular orbits ({\it dotted line}).\label{fig:tranyield}}
\end{figure}

Orbital eccentricity results in an enhanced probability for a planet
to transit and potentially a reduction in the transit detectability.
The overall yield from a transit survey is given by $N_{\rm
det}\propto{\rm Prob}_{{\rm T}}\times N_{\rm obj}$.  The results from
this study can be used to scale the overall yield from a transit
survey based on assuming all planets are on circular orbits for an
assumed distribution of orbital eccentricity.  The enhanced
probability for a planet to transit, ${\rm Prob}_{{\rm T}}$ with a
distribution of orbital eccentricity scaled to the circular orbit case
is given by Equation~\ref{eq:probtrane}.  The reduced number of
transiting planets detectable $N_{\rm obj}$ scaled to the circular
orbit case for an ideal transit survey where the photometric noise is
white and dominated by Poisson error is given in \S~\ref{sec:disc}.
Multiplying these two factors provides the overall yield of an ideal
transit survey scaled to assuming all planets are on circular orbits.
Figure~\ref{fig:tranyield} show that these two opposing effects nearly
cancel over the parameters that characterize the eccentricity
distribution.  Thus, for an idealized transit survey with the
currently observed orbital eccentricity distribution, the overall
yield will be 4\% greater than assuming all planets are on circular
orbits.  The result for an idealized transit survey with a Rayleigh
distribution of orbital eccentricities gives a similar enhancement
(4\%).

In ground-based transit surveys, correlated measurements limit transit
detectability \citep{PON06} and intrinsic variability of the star will
limit the detectability for Earth-sized planets for the Kepler mission
\citep{JEN02}.  In both cases the transit detectability is independent
of the transit duration, in which case $N_{\rm obj}$ is independent of
the orbital eccentricity.  For these cases, the transit survey will
have higher returns by a factor of $\langle {\rm Prob}_{{\rm
T}e}\rangle$ (Equation~\ref{eq:probtrane}) than estimated by assuming
all planets are on circular orbits.  However, the reduced planet yield
in a transit survey due to intrinsic stellar variability or correlated
measurements must be properly accounted for.  If every dwarf star has
an Earth-sized planet orbiting in the habitable zone, then assuming
circular orbits, the Kepler mission expects to detect 100 Earth-sized
planets in the habitable zone \citep{BORU04}.  The work presented here
indicates Kepler will have $\langle {\rm Prob}_{{\rm T}e}\rangle\sim
25\%$ higher yield if Earth-sized planets in the habitable zone have a
planet eccentricity distribution similar to the currently known sample
of giant planets from radial velocity surveys.  However, if
Earth-sized planets with e$\sim$0.9 are as common as e$\sim$0 then,
the yield of Earth-sized planets could be 80\% higher from the Kepler
mission assuming the high-e, short transit duration planets are still
detectable.

The dependence of the transit survey yield on the uncertain underlying
orbital eccentricity distribution implies an uncertainty in measuring
the frequency of terrestrial planets in the habitable zone \citep[a
major goal of the Kepler mission][]{BORU04} .  An analysis of the
transit yield from a transit survey that assumes all planets are on
circular orbits will overestimate the frequency of habitable planets
if high eccentricities are common and not taken into account. 
In practice a variety of
noise regimes affect a transit survey and accurate yields necessitate
an accurate understanding of the photometric noise, stellar sample,
and underlying eccentricity distribution \citep{BUR06,GOU06,FRE07}.

\acknowledgements
This paper benefited from discussions with Scott Gaudi, Will Clarkson,
Peter McCullough, and Eric Ford.  This work is funded by NASA Origins
grant NNG06GG92G.

\end{document}